\documentclass[aps,twocolumn,superscriptaddress]{revtex4}
\usepackage{amsmath,amssymb}
\usepackage{graphics,graphicx}
\usepackage{subfigure}
\usepackage{dcolumn,bm}
\usepackage{psfrag}
\usepackage[dvipsnames]{xcolor}
\usepackage[utf8]{inputenc}
\usepackage{multirow}
\usepackage{braket}
\usepackage{orcidlink}
\usepackage{hyperref}

\newcommand{\vc}
{\affiliation{Vidyasagar College, 39, Sankar Ghosh lane, Kolkata 700006, India.}}
\newcommand{\be}
{\begin{equation}}
	\newcommand{\ee}
	{\end{equation}}

\begin{document}
\title{Does Partial Consumption Help Foraging?}
\author{Md Aquib Molla \orcidlink{0000-0003-0416-1349}}
\vc
\author{Sanchari Goswami \orcidlink{0000-0002-4222-5123}}
\vc
\date{}

\begin{abstract}
In this work, we consider partial consumption of food by a forager in presence of a threshold energy level. The forager considered here can survive for $S$ steps without food, namely the survival time. The threshold limits the consumption of food in such a way that, the forager will only consume food, whenever its energy is below the threshold $k$. Due to partial consumption of food, a site containing food may not always be fully depleted, which in turn helps in increasing the lifetime of the forager. It has been observed that, in our case, the lifetime always increases with $k/S$, although there is a transition threshold $k^*$ below which the increase of lifetime is rapid and above is slow. The transition threshold $k^* \sim \sqrt{S}$. The lifetime $\tau$ shows a power law behavior as $\tau \sim S^{\beta}$. For $k/S=0$, the value of $\beta$ is $1.331$, it then jumps above $2$ and decreases gradually to $1.833$ with increasing $k/S$. Other important quantities like number of revisits to a site, food statistics etc. have been studied and some interesting scaling behaviors are observed. Although, survival for each time unit requires, on an average, the consumption of one unit of food, the food consumption is not the only factor to control the lifetime of the forager. It has been observed that, the strategy in terms of threshold energy and partial consumption affects the lifetime in a positive way.
The collection of sites either fully or partially depleted of food after the death of the forager shows a crossover behavior for $k/S \sim 0.5$.  
\end{abstract}
\maketitle

\section{Introduction}
	Is it necessary for a living organism to always eat whenever it comes across some food in a fear, that, later it may not find some food? The answer may be a `yes' or `no', but, what we observe in reality is that, it does not. Searching for resources, i.e., Foraging is a natural tendency for all living beings. Among all the resources, food is the  most important condition for life. 
    
	In general, for any forager, the elemental attempt is to develop a strategy to maximize its lifetime in terms of food consumption. A concept closely related to this is optimal foraging theory \cite{Charnov1976}. The strategies depend on two main factors: first, searching the food and then, its consumption. The forager always tries to move to a potentially richer domain. These strategic foraging problems are also limited to study the food consumption and lifetime of a forager, but are applicable in several other situations, viz., multiarm bandit problem \cite{Robbins, Gittins}, Feynman’s restaurant problem \cite{Gottlieb}, human memory \cite{Hills, Abbott}, Kolkata Paise Restaurant problem \cite{Chakrabarti} etc., a few of the problems although do not consider the depletion of resources. 
	
	 The environment through which the forager is moving in search of food, may have heterogeneous or uniform food distribution. The forager may survive for a certain number of steps after consuming food from one site. Several possibilities for the motion and action of the forager have been considered in the recent past. This includes the studies of the statistical properties of starving random walks and inter-visit time statistics \cite{Regnier2024, RegnierReview2024}. The simplest case is of course when the forager performs diffusive motion, which can be described within the general framework of random walk theory \cite{Redner, Hughes, Weiss}. In such cases, both the options of taking food whenever it encounters a site containing food \cite{Redner2014} or only when its energy falls below a certain threshold value \cite{Redner2018}
 have been considered. Studies have also been made with different approaches in \cite{Perman, Benjamini, Antal} with consideration of bias towards one side for a one dimensional  walker/forager only when food is consumed. In certain cases, the forager may have some sort of intelligence/smartness and its walk is not always a diffusive one when it is close to the food \cite{Redner2022}.  Also, other factors like  greed may control the movement \cite{Redner2017}.  Related modifications of foraging dynamics involving intermittent resting have also been explored recently \cite{Molla2026}.
	
 In the model proposed in the present work, we consider the possibility of
 partial food consumption which may be beneficial for the forager in two senses:
 the food gets reserved for longer time and the forager lives longer. 
 Although the effect of a threshold level was earlier studied in \cite{Redner2017} for a normal forager, here we study it's effect together with partial consumption on various relevant quantities. The main objective is to check whether partial consumption of resource (here food) which lead to reservation of resource (food) for future use, is beneficial for the forager. In this connection, lifetime of the forager will be studied. Also, consumption depending upon the threshold level, which may be thought of as a strategy, will have some effect in the use of resource. That will also be studied. The energy of the forager, the revisitation statistics, consumed food distribution etc. are other important quantities to be studied here. 
 Section \ref{desc} provides a detailed description of the model. The results are mentioned in section \ref{result}  and the discussions are presented in section \ref{discussion}.

\section{Model Description} \label{desc}

Forager Dynamics models the movement and energy consumption of a forager navigating through a spatial environment, where resources (in this case, food) are distributed across a lattice. The forager has to move as well as to survive. These are governed by a few key parameters, as follows:
\begin{enumerate}
    \item Starving time $S$ : This parameter defines the number of steps the forager can take after consuming food before it starves to death. The forager starts from a designated lattice site, consumes food, and becomes \textit{satiated} (full). It then can take step in a random direction  in the lattice, with its survival depending on the availability of food.
    \item  Threshold energy $k$ : The importance of this parameter is that, whenever the energy of the forager falls below the threshold value $k$, it consumes food, provided it has reached a site containing food. The amount of food consumed depends on the available food there. 
\end{enumerate}
It is worth mentioning that the maximum amount of food the forager can take is just to replenish its energy to become full. Any excess food is left at the site, which may be available to potentially sustain it for a future move. If the food is not sufficient to make the forager full, it will consume the entire available portion and move again.

  We consider a $1d$ lattice where the food distribution is homogeneous and each site initially contains $S$ units of food. The energy of the forager at any time $t$ is termed as $E(t)$. This energy is actually the amount of food with the forager at $t$. The forager starts at the origin consuming the total food there with energy $E(0)=S$. 
  At time $t$, the state of the system is defined by the position $x(t)$, the energy $E(t)$ of the forager. The food distribution is given by $f(x,t)$.

  The dynamics proceed in discrete time steps according to the following rules:

\begin{itemize}
    \item Movement: The forager moves randomly to a nearest-neighbor site,
    \[
    x(t+1) = x(t) \pm 1.
    \]
    \item Energy loss: Each step of movement reduces the energy of the forager by one unit,
    \[
    E(t) \to E(t) - 1.
    \]
    \item Starvation: If $E(t)=0$, the forager dies.
    \item Consumption rule: If $E(t) \leq k$ and the site contains food ($f(x,t)>0$), the forager consumes an amount of food $\Delta$ according to the rule
    \[
    \Delta = \min(S - E(t), f(x,t)).
    \]
    The energy and food distribution are updated as:
    \[
    E(t) \to E(t) + \Delta, \quad f(x,t) \to f(x,t) - \Delta.
    \]
\end{itemize}

It needs to be mentioned here that two extreme cases of the threshold energy are $k=0$ and $k=S-1$.
The first, i.e., $k=0$ indicates a maximally frugal forager \cite{Redner2018} which can only take food when it's energy fuel is totally exhausted. The second case is a tale of a normal forager which takes food, if available, after each hop, although here food may be consumed either partially or fully. The lifetime of the forager, denoted as $\tau$, is defined as the total number of time steps the forager can survive before it dies of hunger. The dynamics of the system depend on both the forager’s ability to manage its energy and the distribution of food across the lattice.

To illustrate the behavior of the forager, consider a 1-dimensional lattice on which the forager moves. To illustrate, we take an example where $S = 4$ and $k=2$ for the forager. 
\begin{itemize}
    \item Initially, at time $t = 0$, the forager starts at $x = 0$, consumes all the food at that site, and becomes full. 
    \item At $t = 1$, the forager randomly moves, say, to $x = 1$. Its energy is decreased by $1$ unit, as it has already taken one step.  
     At this point, it can take  $S-1 = 3$  more steps without consuming food. 
    \item At $t=2$, the forager randomly moves to $x=2$. The energy remaining is $2$. Therefore, $2$ units of food will be consumed by the forager to become full and $2$ units will be left at $x=2$. 
    \item At $t=3$, suppose that the forager moves to $x=3$, its energy is $3$ units and therefore no food will be consumed.
    \item At $t=4$, if the forager moves to $2$ its energy is $2$ and it will consume the amount of food remaining at $x=2$ and will become full again.
    \item It then moves to $x=1$ and $x=0$ at $t=5$ and $t=6$ respectively. Its energy is decreased to $2$. Now there is no food at $x=0$.
    \item At $t=7$, suppose it moves to $x=-1$. There it will consume $3$ food units and $1$ food unit will be left at $x=-1$.
    \item In the next time steps from $t=8,9$, if the forager moves to $-2, -1$, it will consume one food unit at $x=-1$ Its energy is $3$ units.
    \item For $t=10-12$, it moves to $0, -1, 0$ and will die..  
\end{itemize}
 In this way, the forager moves. For this particular case, the lifetime of the forager is $12$. 
For the last time step $t=12$, if the forager moves to $x=-2$, a food containing site, it will consume the food and can save its life. In that case, if we consider the above example, the lifetime of the forager can be greater than $12$. However, in reality, 
if the forager moves to $x=-2$, a food containing site, it will still die, as its energy is exhausted on reaching the site and it will not even be able to consume the food. To make things simpler, we have not taken this into account in the present work. We have checked that `no food consumption at the last step' when the energy of the forager reaches to $0$ due to movement, do not change the nature of quantities studied much. This will be discussed in the following sections alongwith the results for the original model.

\section{Results}\label{result}

\subsection{Study of Average lifetime}
\begin{figure}[h!]
	\begin{center}
		\includegraphics[angle=-90, trim = 0 0 0 0, clip = true, width=0.99\linewidth]{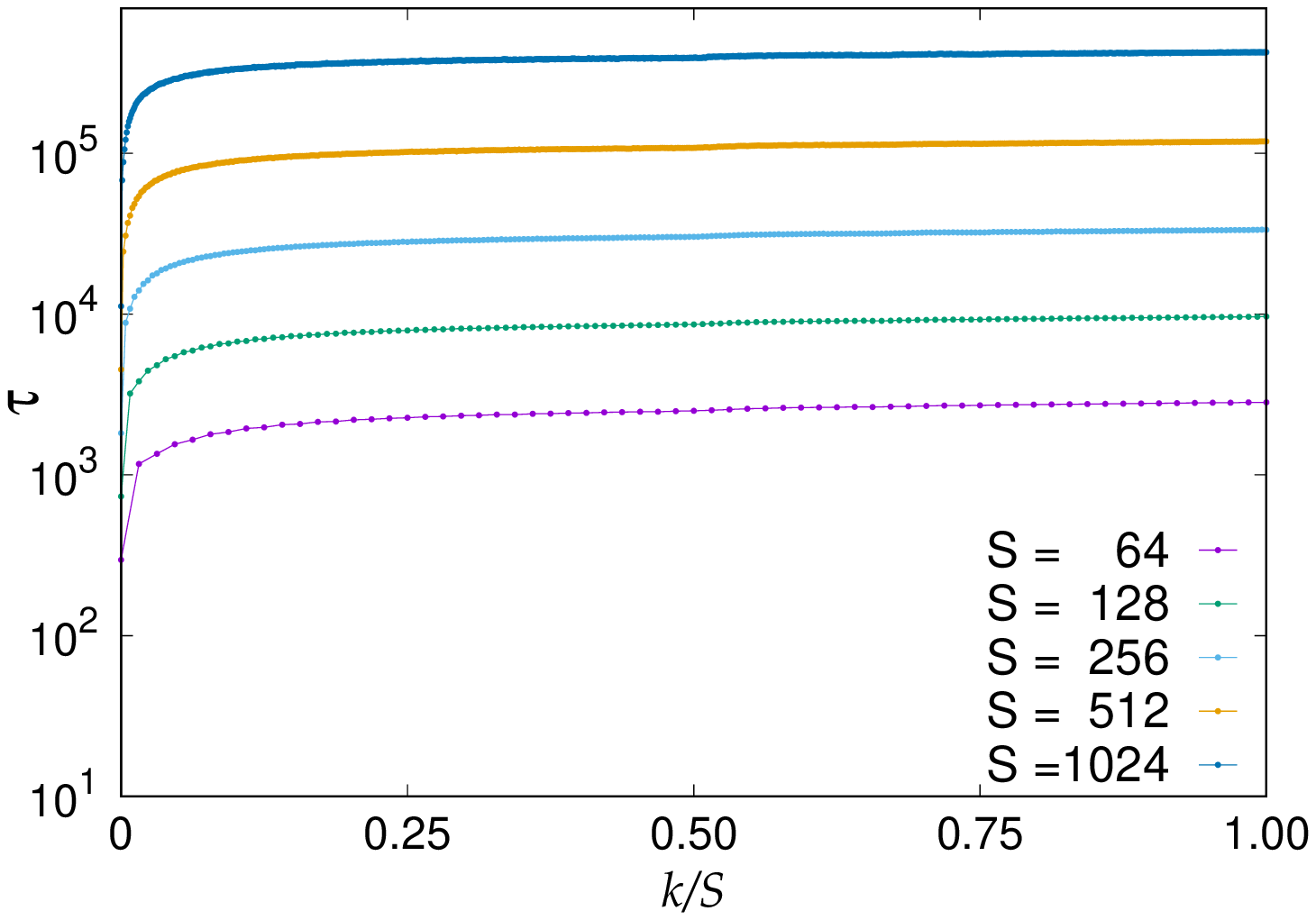}
	\caption{Study of lifetime $\tau$ against threshold energy fraction $k/S$ with $S$ as a parameter. It has been observed that in case of partial food consumption, the lifetime never decrease beyond a certain $k^*/S$.}
		\label{ComparisonModel1Model2}
	\end{center}
\end{figure}

In the $1d$ case, we examine the average lifetime $\tau$ of the forager as a function of $k/S$. Here $k/S$ may be called a threshold energy fraction. This has been done keeping $S$ as a parameter and by taking $10^5$ realizations. Earlier in \cite{Redner2018}, it was observed that the average forager lifetime is maximum at an optimal threshold. In our work, we found that as we increase $k/S$ from a very low value, increase of $\tau$ is rapid at the beginning. Beyond a certain $k^*/S$, $\tau$ increases slowly and never decreases thereafter. This threshold $k^*$ may be called a transition threshold and $k^*/S$ may be called the transition threshold fraction, as below it the change of lifetime is rapid and above is very slow. This is shown in Fig. \ref{ComparisonModel1Model2}. 

The results of Fig. \ref{ComparisonModel1Model2} are interesting in the sense that beyond a certain $k^*/S$, the lifetime never decrease for our model with partial consumption. For very low threshold value, $\tau$ is very small because the forager only eats when all its consumed food is exhausted on an average. Thus, there may be chances of quick death. However, for a fixed $S$, increasing $k$ beyond $k^*$ together with partial consumption always ensures a favorable condition for the survival of the forager.  


\begin{figure}[h!]
	\begin{center}
		\includegraphics[angle=-90, trim = 0 0 0 0, clip = true, width=0.99\linewidth]{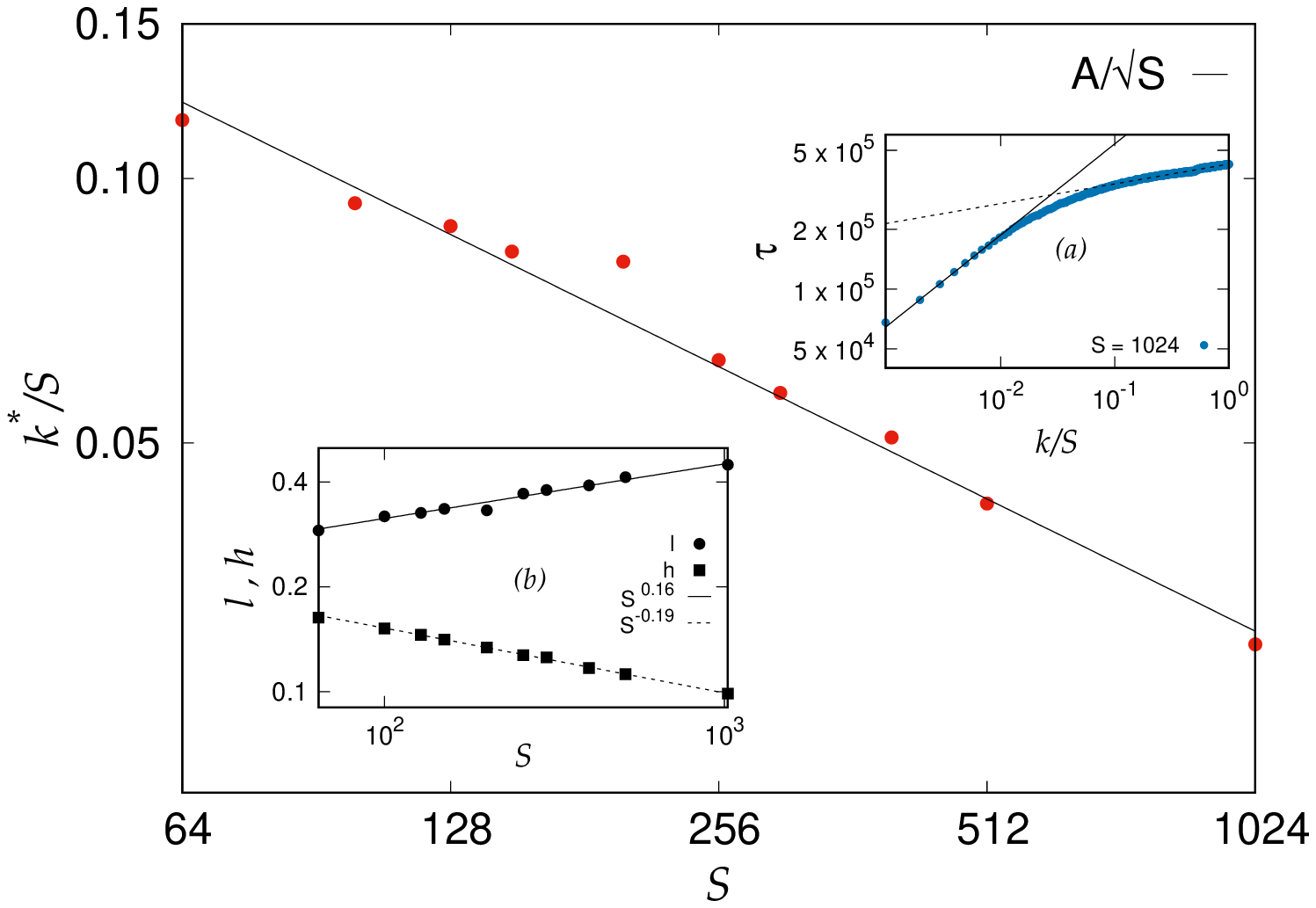}
		\caption{Plot of $k^*/S$ as a function of starvation time $S$, showing the scaling $k^* \sim A\sqrt{S}$ with $A = 0.978 \pm 0.008$
		The top-right inset (a) shows the variation of the average lifetime $\tau$ with $k/S$ for $S = 1024$, illustrating how $k^*/S$ is determined. The fitting lines below and above $k^*/S$ show exponents $l$ and $h$ respectively. The other left bottom inset (b) shows the variation of the fitting exponents $l$ and $h$ with $S$.}
		\label{Comparison_Model1Model2}
	\end{center}
\end{figure}



For a better understanding of how lifetime is affected with the threshold energy fraction $k/S$ and to find the value of transition threshold $k^*/S$, let us take a look at Fig. \ref{Comparison_Model1Model2}. In the inset (a), the variation of $\tau$ with $k/S$ is shown on a log-log scale for $S=1024$. It is clear that the variation fits a power law but below (shown by solid line) and above (shown by dashed line) a certain $k^*/S$ differently as shown in Fig. \ref{Comparison_Model1Model2} (a). Thus
\begin{align}
	\tau = A \left( \frac{k}{S} \right)^{l} \hspace{0.1cm}; \hspace{0.2cm} k < k^* \nonumber \\
	\hspace{0.3cm} = A \left( \frac{k}{S} \right)^{h}\hspace{0.1cm} ; \hspace{0.2cm} k > k^*
\end{align} 

Straight lines are fitted for low $k/S$ and for high $k/S$. The straight lines intersect each other and from that, the transition threshold $k^*/S$ is determined. For example, in Fig. \ref{Comparison_Model1Model2}(a), for the range $0.0009-0.01$, the line on a log scale is fitted as $(k/S)^l$, whereas,  for the range $0.09-1$, it is fitted with $(k/S)^h$. The corresponding $k^*/S=0.0295$.\\
In Fig. \ref{Comparison_Model1Model2}(b), the variation of $l$ and $h$ with $S$ are shown. It has been observed that, $l$ increases while $h$ decreases with increase in $S$, both as power laws.
 
In the main part of Fig. \ref{Comparison_Model1Model2}, variation of $k^*/S$ is shown as a function of $S$. From the nature of the plot, we have found that
\begin{align}
	\frac{k^*}{S} &\sim \frac{A}{\sqrt{S}} \nonumber \\
	k^* &\sim A\sqrt{S}
	\label{k*}
\end{align}
where $A \simeq 1$.
This can be explained as follows: 
In the case of partial foraging, eating does not permanently remove food from the environment always. The forager consumes only the amount of food needed to replenish its energy; any remaining food is available for its future visits. Therefore, as $k$ increases, the forager is allowed to eat earlier. This reduces the food availability, but partially. 

For a random walk, the number of distinct sites visited up to time $t$ scales as $\sqrt{8t/\pi}$, i.e. the size of the desert grows with time as $t^{1/2}$. For a forager with starvation capacity $S$, the size of the desert is of the order of $S^{1/2}$. If the threshold energy $k < S^{1/2}$ then the forager most likely will starve inside the desert. However, when $k > S^{1/2}$ the forager may encounter food at the boundary. Equating these two length scales yields $k^* \sim S^{1/2}$. \\

It is worth mentioning that in $2$d, the lifetime of the forager behaves in a same way as in $1$d, i.e., beyond a certain transition threshold lifetime still increases, but slowly. The scaling behavior of $k^*$ remains unchanged for $2$d, i.e., $k^* \sim S^{1/2}$.

\begin{figure}[h!]
	\begin{center}
		\includegraphics[angle=-90, trim = 0 0 0 330, clip = true, width=0.99\linewidth]{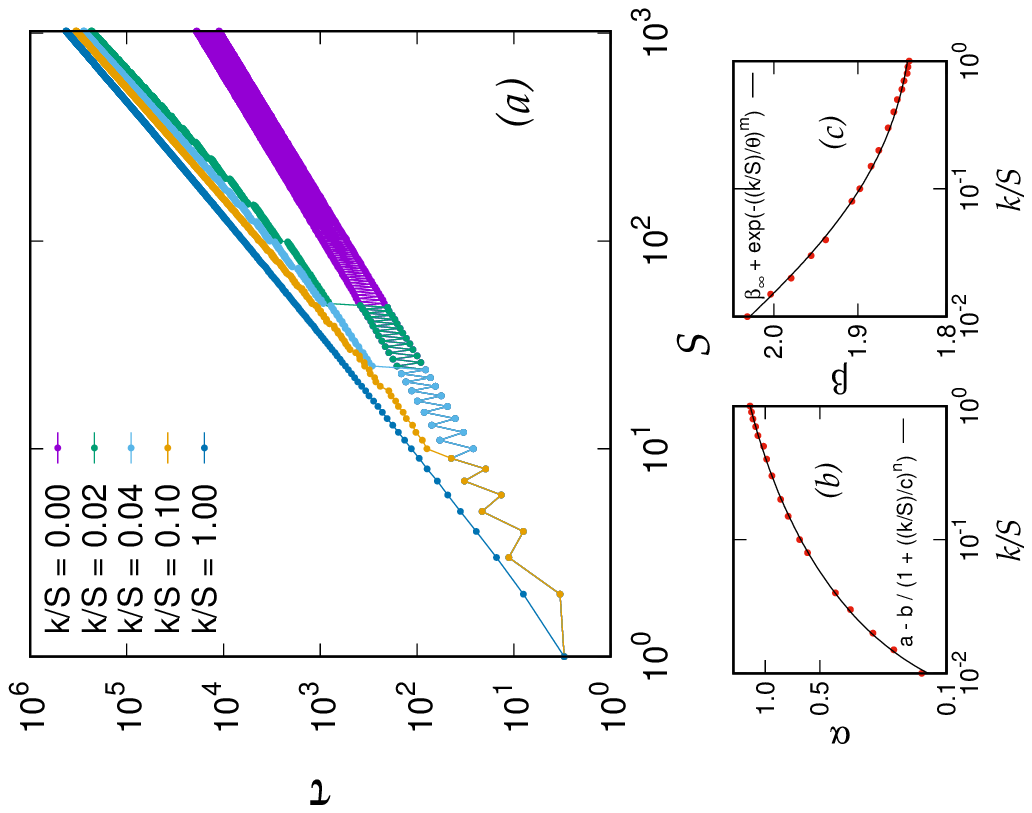}
		\caption{Lifetime $\tau$ against starving time $S$ with varying $k/S$ values. For $k/S=0.00$, the exponent $\beta \simeq 1.331$. 
In asymptotic region of $S$, $\tau = \alpha S^{\beta}$. In the inset, (b) the variation of $\alpha$ and (c) the variation of $\beta$ have been shown against $k/S$; $\alpha = a - \frac{b}{1+\left(\frac{(k/S)}{c}\right)^{n}}$ and $\beta = \beta_{\infty} + \exp\left(-\left(\frac{(k/S)}{\theta}\right)^{m}\right)$. For $k/S$ slightly above $0$, the exponent $\beta$ jumps above $2$ and then finally reaches a value $\beta_{\infty} \approx 1.833$ for $k/S \to 1$. }
		\label{TaveModel1Model2}
	\end{center}
\end{figure}

Next, we are going to study the response of lifetime $\tau$ with $S$ keeping $k/S$ a parameter. From Fig. \ref{TaveModel1Model2}, it has been observed that for moderate to high $S$, lifetime $\tau$ scales with $S$ according to a power law:
\begin{equation}
    \tau = \alpha S^{\beta}
\end{equation}
where $\alpha$ and $\beta$ both depend on the threshold energy fraction $k/S$. As $S$ increases, the forager can take more steps by consuming food each time. Thus, it can sustain itself for a longer duration without the need to consume food. This helps prolonging its foraging lifetime $\tau$. This is shown in Fig. \ref{TaveModel1Model2}. For $k/S=0.00$, the exponent $\beta = 1.331 \pm 0.003$. 
This is quite close to the results as obtained in \cite{Redner2018} for $k=0$. In Fig. \ref{TaveModel1Model2}(b) and (c) the variation of $\alpha$ and $\beta$ are shown. For $k/S$ slightly above $0$, the exponent $\beta$ jumps above $2$ and then finally reaches a value $\beta_{\infty}$ for $k/S \to 1$. It has been observed that the functional form of $\alpha$ and $\beta$ are as follows $\alpha = a - \frac{b}{1 + \left( \frac{(k/S)}{c} \right)^{n}}$ and $\beta = \beta_{\infty} + \exp\left(- \left( \frac{(k/S)}{\theta} \right)^{m}  \right)$, with $a = 2.001 \pm 0.251$, $b = 2.412 \pm 0.409$, $c = 0.185 \pm 0.042$, $n = 0.428 \pm 0.079$ $\beta_{\infty} = 1.833 \pm 0.002$, $\theta \simeq 0.0012$ and $m = 0.226 \pm 0.006$.

We have mentioned earlier that a real forager, that dies when all its energy is exhausted on reaching a site, even if the site contains food. We have checked that in that, for $k/S>0.00$, $\tau$ behaves in the same way with $S$, i.e., exponents $\beta$ are same, as in the case of our model as above. Here, $\beta_{\infty}=1.837\pm 0.001$. Only for $k/S=0.00$, $\tau \sim S$, the exponent $\beta$ being $1$. This is because as threshold $k=0$, the forager exhausts all its energy but unable to take the food upon availability at the last step. Thus, its lifetime will be equal to the amount of energy exhausted, i.e., energy taken $S$.


We would now present some mathematical argument for the exponents in the extreme limits of $k/S$ for our original model. To make a completely empty site the walker requires to visit the site multiple times. Considering the average amount of consumption on a particular site $S-k$ and number of visits to create an empty site $N_d$, we have
\begin{equation}
    N_d(S-k) \sim S \quad \text{or,} \quad N_d \simeq \frac{S}{S-k}.
\end{equation}

Now, the desert radius $R(t) \sim \sqrt{t/N_d}$. Starvation occurs when this radius exceeds the diffusive reach. Implies, $$R(t) \sim \sqrt{k}$$. \\
Therefore, 
\begin{equation}
    \sqrt{\tau/N_d} \sim \sqrt{k} \quad \text{or,} \quad \tau \sim N_d k 
\end{equation}
which implies,
\begin{equation}
    \tau \simeq \frac{S}{S-k} k.
\end{equation}
Thus, in the limit of large $k$, when $k \sim S-1$, we obtain
\begin{equation}
\tau \sim S^2.
\end{equation}

When $k \ll \sqrt{S}$, the walker typically explores a region much larger than the length scale introduced by the threshold before consuming any food. As a result, the threshold does not significantly restrict the walker’s motion inside the depleted region (desert). Starvation is therefore governed primarily by the walker’s ability to diffusively escape the desert, rather than by the threshold rule. Consequently, the dynamics reduce to those of the classical starvation random walk, yielding the lifetime scaling
\begin{equation}
    \tau \sim S^{4/3}.
\end{equation}

It is to be mentioned that the lifetime and the related threshold and exponents remain unchanged, if the food distribution is heterogeneous.

\subsection{Distribution of revisits $N(x)$}
\begin{figure}[h!]
	\begin{center}
		\includegraphics[angle=-90, trim = 0 0 0 0, clip = true, width=0.99\linewidth]{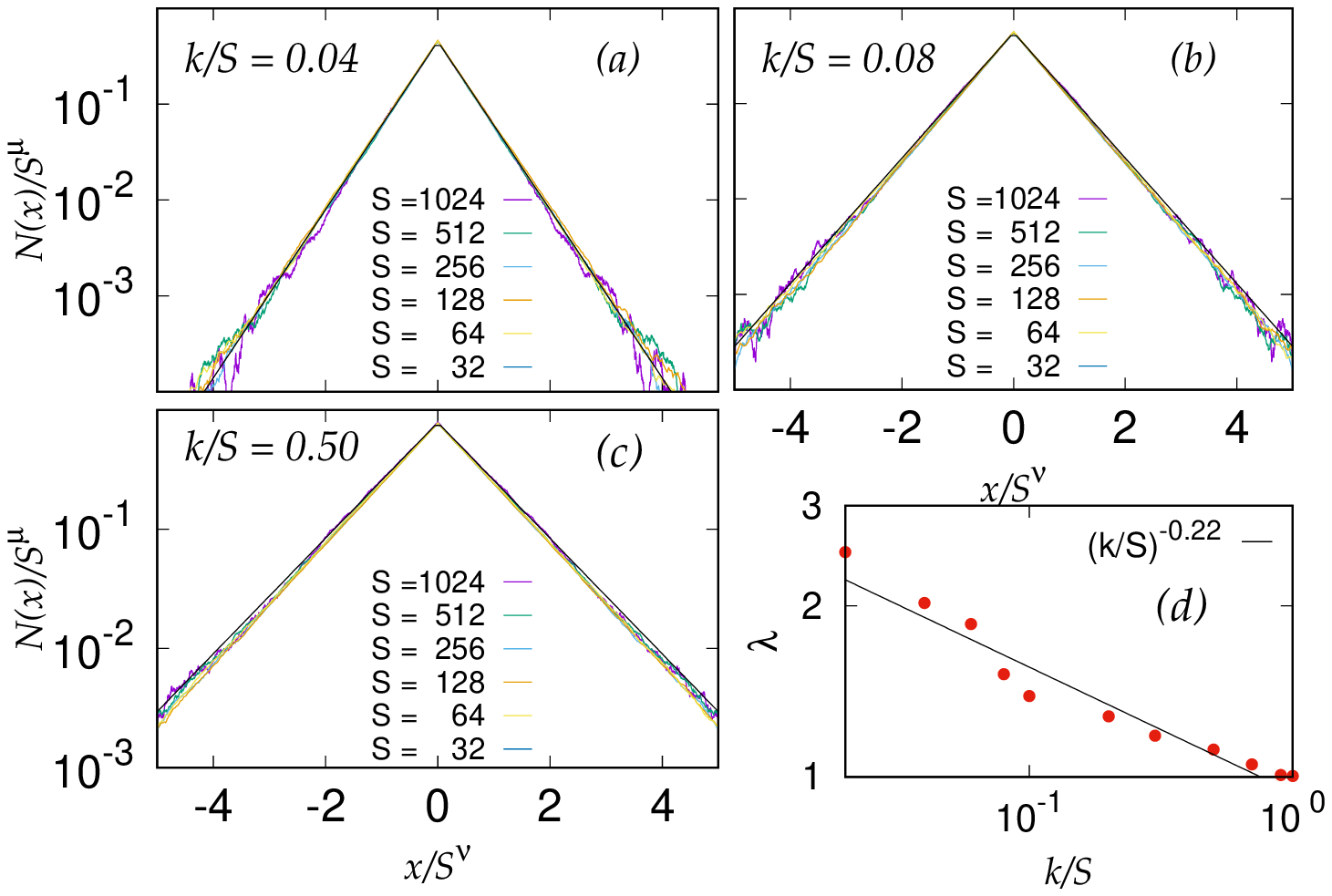}
		\caption{Plot of scaled number of revisits $N(x)/S^{\mu}$ against $x/S^{\nu}$ with $S = 32, 64, 128, 256, 512$ and $1024$. In the subplot $(a)$, $(b)$ and $(c)$; $k/S = 0.04, 0.08$ and $0.50$ respectively. The fitted function is $N(x)/S^{\mu} = N_0 \exp{(-\lambda |x/S^{\nu}|)}$. In $(d)$ the variation of $\lambda$ against $k/S$ has shown with the exponent = $-0.22 \pm 0.02$.} 
		\label{N_vs_x_collage}
	\end{center}
\end{figure}

We computed the distribution of $N(x)$, the number of times the forager visits a given site $x$. In Fig. \ref{N_vs_x_collage}(a) - (c), the data collapse for the same has been shown for different $S$ with scaled variables $x/S^{\nu}$ and $N(x)/S^{\mu}$. It is clear that the distributions exhibit exponential decay. As $S$ becomes smaller, fewer steps are available to the forager. This observation is consistent with the power-law dependence of $\tau$ on $S$, which implies that a higher starving time $S$ results in a longer lifetime. The values of $\mu$ and $\nu$ are shown in Table \ref{Table 1}.


\begin{table}[]
	\centering
	\begin{tabular}{|c|c|c|}
		\hline 
		\hspace{2pt} $k/S$ \hspace{2pt} & $\nu$ & $\mu$\\ 
		\hline
		\hline
		0.02 & \hspace{5pt} 0.975 \hspace{5pt} & \hspace{5pt} 0.975 \hspace{5pt}\\
		0.04 & 0.960 & 0.965\\
		0.06 & 0.955 & 0.960\\
		0.08 & 0.930 & 0.945\\
		0.10 & 0.920 & 0.930\\
		0.20 & 0.915 & 0.920\\
		0.30 & 0.905 & 0.915\\
		0.50 & 0.900 & 0.910\\
		0.70 & 0.895 & 0.905\\
		0.90 & 0.890 & 0.900\\
		1.00 & 0.890 & 0.900\\
		\hline
	\end{tabular}
	\caption{Approximate values of $\nu$ and $\mu$ from Fig. \ref{N_vs_x_collage} for different $k/S$.}
	\label{Table 1}
\end{table}

\subsection{Energy of Forager}
\begin{figure}[h!]
	\begin{center}
		\includegraphics[angle=-90, trim = 0 0 0 0, clip = true, width=0.99\linewidth]{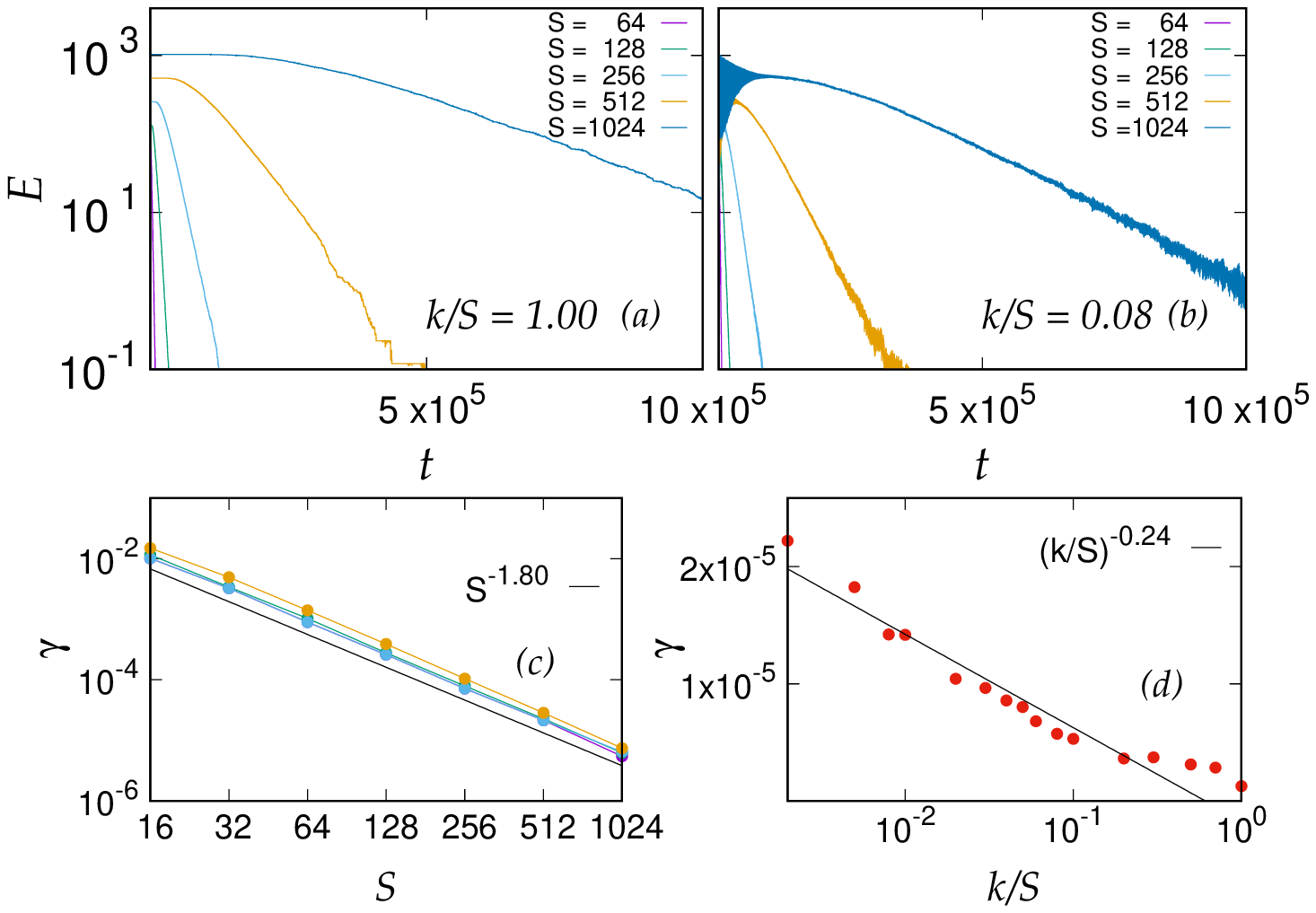}
		\caption{Plot of energy $E$ versus time $t$ for $(a)$ $k/S = 1.0$ and $(b)$ $k/S=0.08$ respectively. The variations are fitted as $E \propto \exp(-\gamma t)$. In $(c)$, the variation of $\gamma$ against $S$ with $k/S = 1.0$ has been shown for different $k/S$. Here, $\gamma \propto S^{-1.80}$, In $(d)$, the variation of $\gamma$ against $k/S$ with $S = 1024$ has been shown and the observed dependence is $\gamma \propto (k/S)^{-0.24}$.}
		\label{1D_E_k_S_inset}
	\end{center}
\end{figure}

It is already known that the energy of the forager increases on food consumption and decreases for taking each step. In Fig. \ref{1D_E_k_S_inset}(a), (b), the plot of $E$ as a function of time $t$ is shown for different $S$ for $k/S=1.00, 0.08$ respectively. It has been observed that the energy variations for large $t$ are exponential, i.e., $E \propto \exp(-{\gamma}t)$. 
The explanation for the observed behavior can be explained as follows : Consider the term $S$, which tells us how long the forager can survive without food. Now, consider a fixed $k/S$. When $S$ is small, $k$ is proportionally small. In case of large $S$, the threshold is also high. In both the cases, the forager will eat as its energy falls below the specified $k$. However, for large $S$, the storage capacity is large for which it can continue for long without food and whenever it eats, the eating proportion is also large and its energy will decay slowly. At the same time, previously visited regions become only partially depleted, so the forager must find food portions at previously visited sites during movement to replenish its energy, which makes the energy decay slower. Here, the decay is not controlled by storage alone; it is closely linked to how the forager explores space for food encounter. As during this exploration the food consumption is partial, there is chances of slow energy decay and therefore longer life. This is shown in Fig. \ref{1D_E_k_S_inset}(c), for $k/S \leqslant 1$, where $\gamma$ is shown as a function of $S$. This fits a decaying power law with exponent close to $1.80$ for $k/S=0.08, 0.2, 0.5, 1.0$. It is also clear from the Figure, that, there is a slight dependence of $\gamma$ on $k/S$. In (d), the variation of the decay rate $\gamma$ as a function of $k/S$ is shown for $S=1024$. The observed dependence of $\gamma$ is given as :  $\gamma \propto (k/S)^{-0.24}$. Increase in $k/S$ for a fixed $S$ means that the forager have permission to eat more frequently, thereby decreasing the value of $\gamma$.

\subsection{Food Consumption}
\begin{figure}[h!]
	\begin{center}
		\includegraphics[angle=-90, trim = 0 0 0 0, clip = true, width=0.99\linewidth]{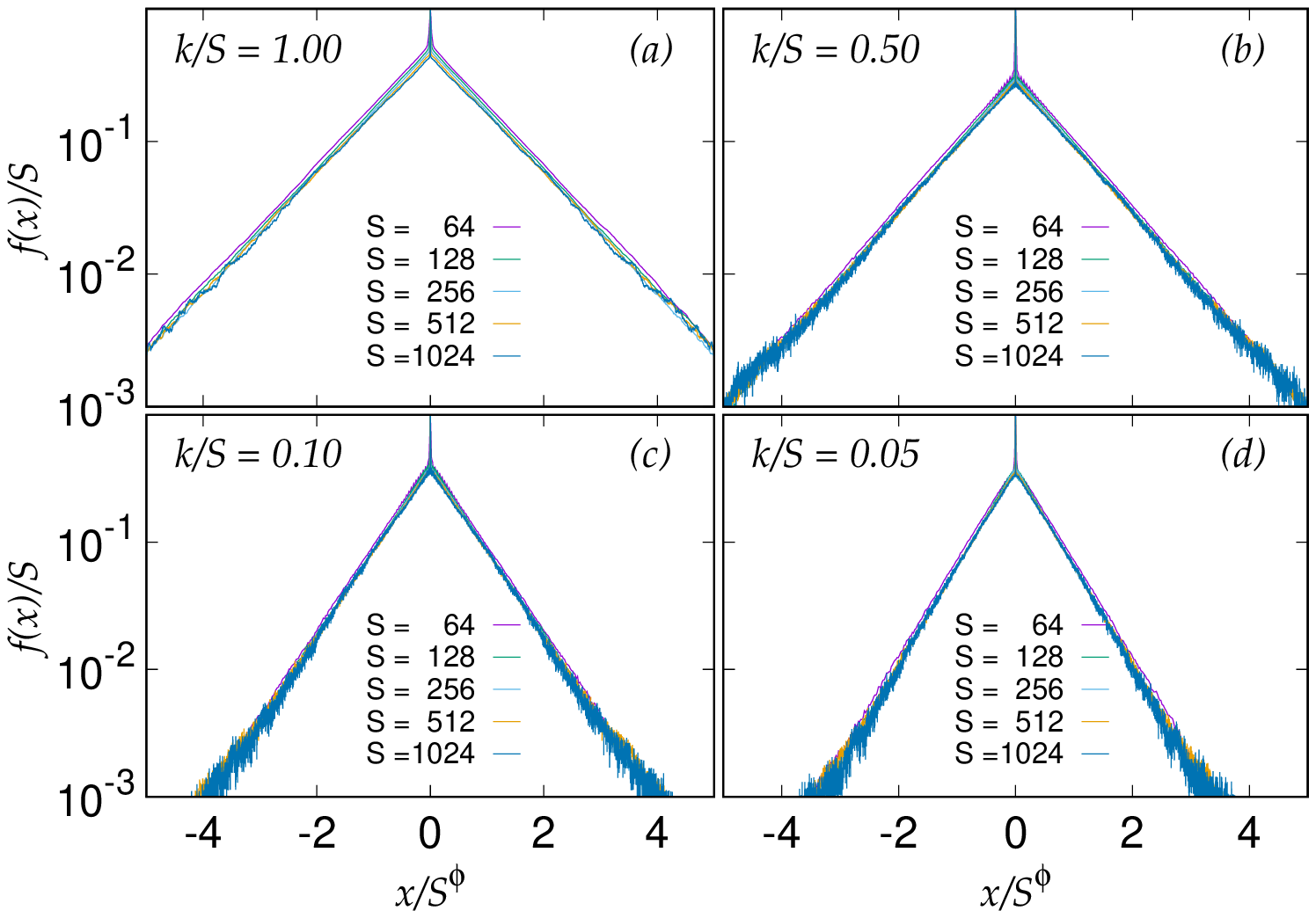}
		\caption{Plot of $f(x)/S$ against $x/S^{\phi}$ with $S =64, 128, 256, 512$ and $1024$ for (a) $k/S = 1.00$, (b) $0.50$, (c) $0.10$ and (d) $0.05$ respectively. The fitted function is $f(x)/S = F_0 \exp{\left(-\frac{|x|}{\xi S^{\phi}}\right)}$. The $\xi$ values for $k/S=1.00, 0.50, 0.10, 0.05$ are $0.938 \pm 0.001, 0.835 \pm 0.001, 0.676 \pm 0.002$ and $0.584 \pm 0.002$ respectively.}
		\label{1D_food_x}
	\end{center}
	\end{figure}

For a typical forager problem, another important quantity to look into is the food consumption. The food consumption governs the survival of the forager. We define $f(x)$ as the total food consumed at lattice site $x$ during the entire lifetime. 
In Fig.~\ref{1D_food_x}(a), (b), (c) and (d), the distribution of consumed food over sites are shown. As $x$ increases, the consumed food decreases exponentially. It has been observed that for a fixed $k/S$ all the curves collapse onto a single curve, when the rescaled profile $f(x)/S$ is plotted against $|x|/S^{\phi}$. 
From the Figure,  the scaling form obtained is as follows:
\begin{equation}
	\frac{f(x)}{S}=F_0\exp\!\left(-\frac{|x|}{\xi\,S^{\phi}}\right),
\end{equation}
where $F_0$ is the amplitude and $1/\xi$ signifies the decay strength. From the dimension, therefore, $\xi$ may be thought of as a decay length. For different $k/S$, the value of $\xi$ and $\phi$ are mentioned in Table \ref{Table 2}. For a few $k/S$, $\xi$ values are mentioned in the caption of Fig. \ref{1D_food_x} too. For smaller values of $k/S$, values of $\xi$ are smaller. Thus, the rescaled profiles become gradually narrower and the forager explores a smaller region of space. The food consumption is therefore also less.

Now, the total consumed food can be expressed as,
\begin{equation}
	f_{\text{tot}}=\sum_x f(x).
\end{equation}

\begin{figure}[h!]
\begin{center}
	\includegraphics[angle=-90, trim = 0 0 0 0, clip = true, width=0.99\linewidth]{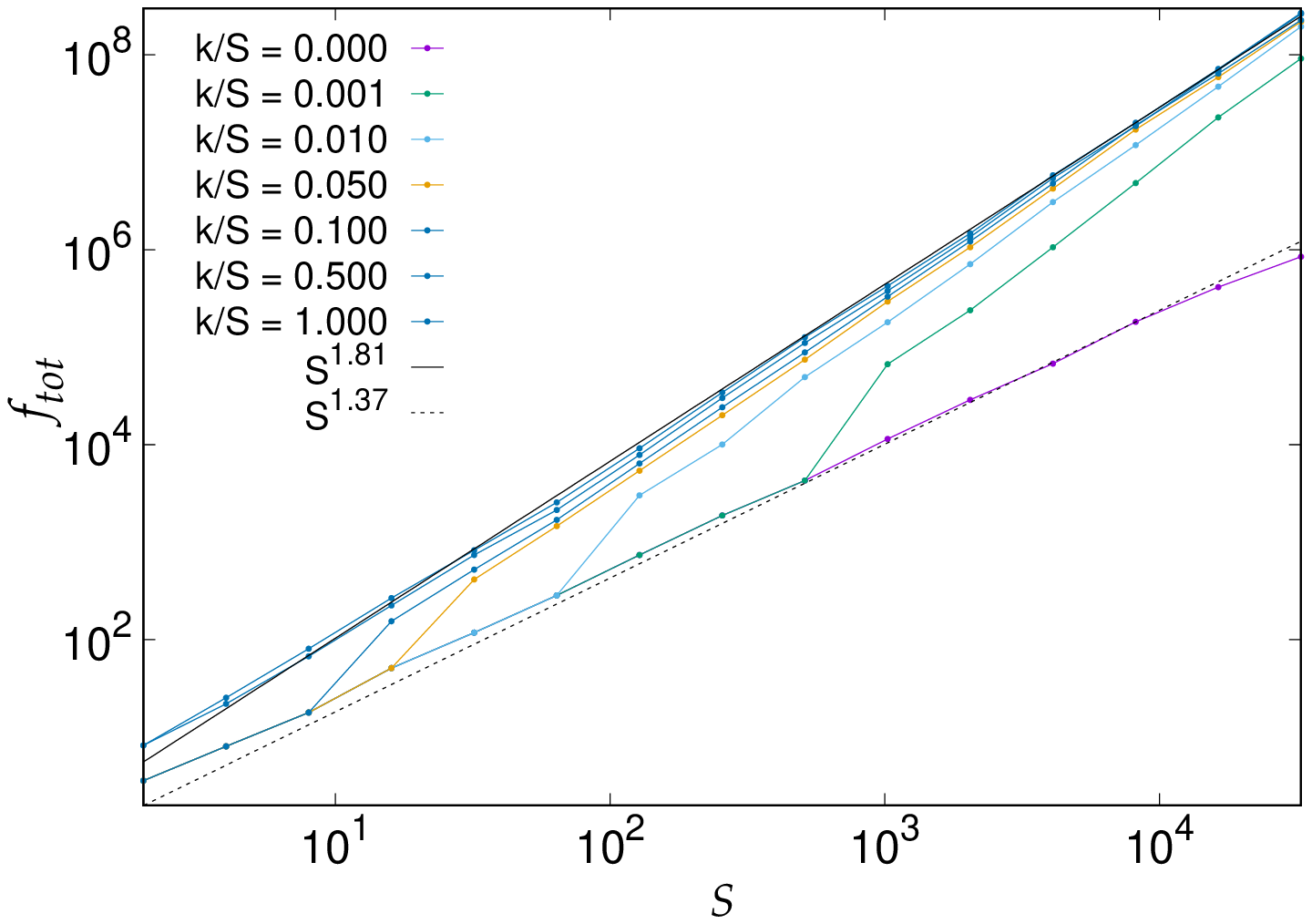}
	\caption{Amount of food eaten $f_{tot}$ versus starvation time $S$ with $0\leqslant k/S \leqslant 1$.}
	\label{ftot}
\end{center}
\end{figure}

The variation of $f_{\text{tot}}$ with starvation time $S$ is shown in Fig.~\ref{ftot}. It has been observed that the scaling relation 
\begin{equation}
    f_{\text{tot}} = \alpha_f S^{\beta_f}
\end{equation}
The fitted parameters $\alpha_f$ and $\beta_f$ are found to match exactly with $\alpha$ and $\beta$ respectively, as in the variation of $\tau$. This implies that $f_{\text{tot}}$ scales identically with $\tau$. This behavior is expected as each time step contributes one unit to the lifetime and survival for each time unit requires, on average, the consumption of one unit of food. 



	\begin{table}[]
\centering
\begin{tabular}{|c|c|c|}
	\hline 
	\hspace{2pt} $k/S$ \hspace{2pt} & $\phi$ & $\xi$\\ 
	\hline
	\hline
	0.03 & \hspace{5pt} 0.970 \hspace{5pt} & \hspace{5pt} 0.426 $\pm 0.002$ \hspace{5pt} \\
	0.05 & \hspace{5pt} 0.940 \hspace{5pt} & \hspace{5pt} 0.584 $\pm 0.002$ \hspace{5pt} \\
	0.07 & \hspace{5pt} 0.930 \hspace{5pt} & \hspace{5pt} 0.609 $\pm 0.002$ \hspace{5pt} \\
	0.10 & \hspace{5pt} 0.930 \hspace{5pt} & \hspace{5pt} 0.676 $\pm 0.002$ \hspace{5pt} \\
	0.20 & \hspace{5pt} 0.920 \hspace{5pt} & \hspace{5pt} 0.712 $\pm 0.002$ \hspace{5pt} \\
	0.30 & \hspace{5pt} 0.910 \hspace{5pt} & \hspace{5pt} 0.778 $\pm 0.001$ \hspace{5pt} \\
	0.50 & \hspace{5pt} 0.900 \hspace{5pt} & \hspace{5pt} 0.835 $\pm 0.001$ \hspace{5pt} \\
	0.70 & \hspace{5pt} 0.900 \hspace{5pt} & \hspace{5pt} 0.865 $\pm 0.001$ \hspace{5pt} \\
	1.00 & \hspace{5pt} 0.890 \hspace{5pt} & \hspace{5pt} 0.938 $\pm 0.001$ \hspace{5pt} \\
	\hline
\end{tabular}
\caption{Approximate values of $\phi$ and $\xi$ from Fig. \ref{1D_food_x} for different $k/S$.}
\label{Table 2}
\end{table}

\begin{figure}[h!]
\begin{center}
	\includegraphics[angle=-90, trim = 0 0 0 0, clip = true, width=0.99\linewidth]{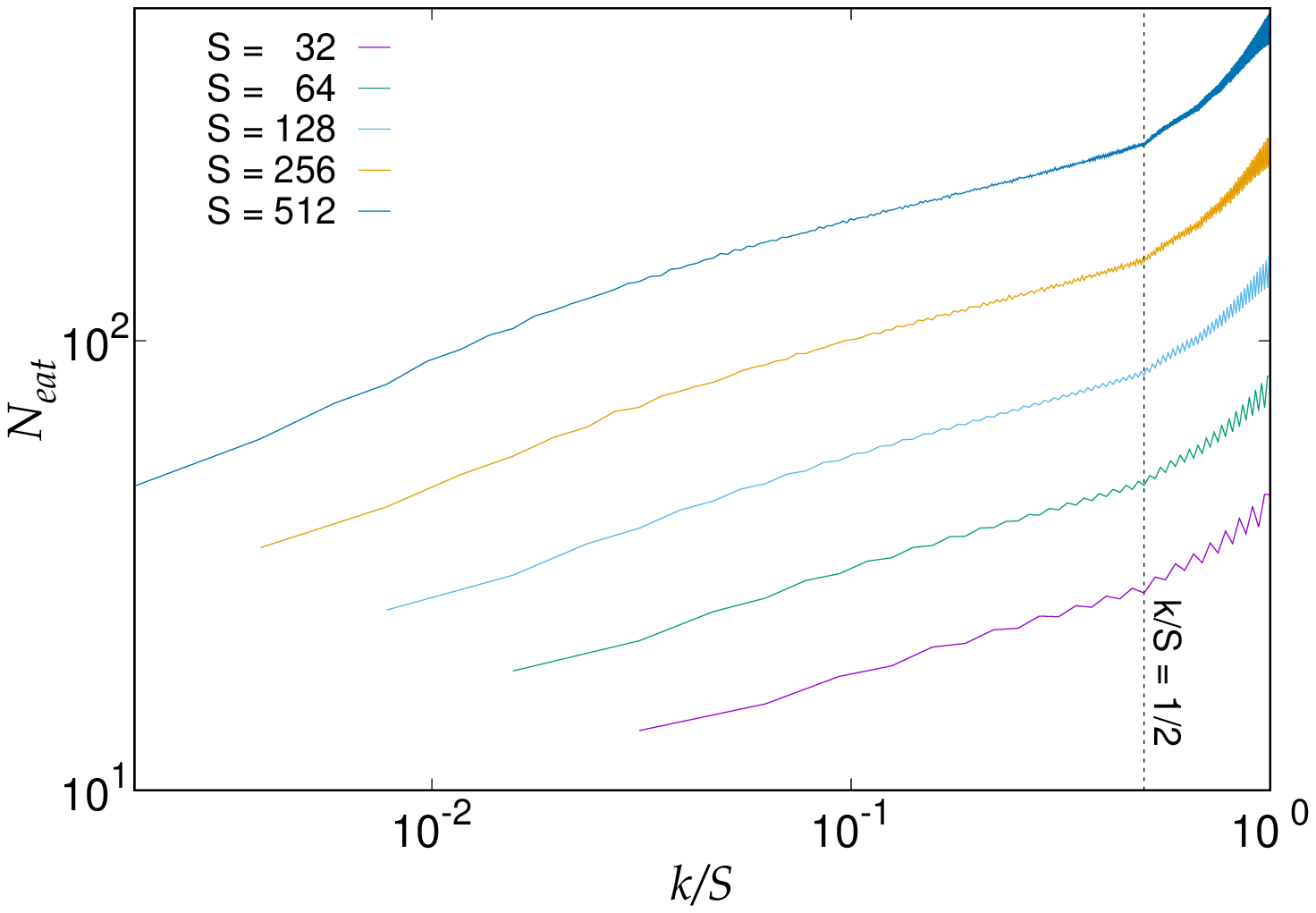}
	\caption{Variation of amount of food eaten $N_{eat}$ against $k/S$ with $S = 32, 64 , 128, 256$ and $512$. Exponent changes near $k/S \sim \frac{1}{2}$.}
	\label{N_eat}
\end{center}
\end{figure}

Now, we are going to present another interesting quantity, $N_{eat}$. Here, $N_{eat}$ is the collection of sites either fully or partially depleted of food after the death of the forager. This means that these sites are visited by the forager and it has eaten food from there. 
In Fig. \ref{N_eat}, we plot, $N_{eat}$, as a function of the threshold fraction $k/S$ for different starvation times $S$. 
The figure shows that $N_{eat}$ increases with both $k/S$ and $S$, indicating that a forager with a higher starvation capacity or an higher consumption threshold explores a larger region of the lattice.

Interestingly, a change in the effective scaling behavior is observed near $k/S \sim 0.5$. For smaller values of $k/S$, the forager delays consumption for which eating becomes less frequent. Therefore, the sites from which the forager eats become almost entirely depleted. This limits the space covered by the forager because of its death within the desert. 
In contrast, for larger $k/S$, the forager eats earlier and maintains a higher energy level, enabling it to travel farther and access newer sites. This crossover therefore reflects a transition from a locally confined search to a more extended exploration strategy. 


It has been already mentioned that, for $k/S \sim 0.5$, a crossover behavior is observed. Below we try to explain the respective behavior:
We know that the food is only consumed by the forager when its energy is below the threshold energy $k$ and the amount of food consumed then is $S-k$. When $k$ is high enough, the average amount of food consumed per hitting $S-k$. In case of small $k$, however, the average consumption is $S -k \approx S$. Therefore, the forager eats almost the entire food in a single visit. The partial consumption here plays an interesting role when the threshold $k$ is sufficiently high. To deplete a site completely, it needs to visit atleast twice. Therefore, in the second visit, the site has enough food to satiate the walker $$k > S - k$$ \\ implies, 
\begin{equation}
    \frac{k}{S} > \frac{1}{2}
\end{equation}
Similarly, for the third time visit $k > 2(S-k)$, implies, $\frac{k}{S} > \frac{2}{3} > \frac{1}{2}$ and so on up to, $n^{\text{th}}$ visit $k > (n-1)(S-k)$ i.e., $\frac{k}{S} > \frac{n-1}{n}$. The effect of partial consumption play significant role after $k/S = \frac{1}{2}$. This clearly explains the change of exponent after $k/S > \frac{1}{2}$ in Fig. \ref{N_eat}.

\section{Discussion}\label{discussion}
In this work, we studied partial food consumption for a forager. In general, the forager considered here can survive for $S$ steps without food; at the same time, it consumes food only when its energy falls below a threshold $k$. In our work, for all the cases of the  forager, whether it is a normal one ($k=S-1$) or maximally frugal($k=0$), the add-on is the partial consumption. A site containing food may not always be fully depleted because of partial consumption. Thus, there is a chance of slow growth of the desert, which in turn helps in increasing the lifetime of the forager. It has been observed that as the threshold fraction $k/S$ increases from a very low value, the lifetime increases abruptly at the beginning, then reaches a high value at some transition threshold $k^*/S$ and never decreases thereafter, although the growth rate becomes low. The transition threshold $k^* \sim \sqrt{S}$. The lifetime $\tau$ when considered with variation of $S$, keeping $k/S$ a parameter, shows a power law behavior as $\tau \sim S^{\beta}$. For $k/S=0$, the value of $\beta$ is $1.331$ which in agreement with the exponent observed in \cite{Redner2018}. With increase in $k/S$, the exponent $\beta$ jumps above $2$ and then decreases gradually to $1.833$. In any case, it can be concluded that the lifetime can be tuned by the threshold fraction $k/S$. Beyond a certain transition threshold $k^*/S$, although the rate of increase of lifetime becomes slow, but it never decreases. Thus, in terms of strategy we can conclude that, by increasing $k/S$, the lifetime can be increased as partial consumption is involved. This is because, when $k/S$ is very high, the food consumption is frequent but in small portions, which makes the growth rate of the desert slow.

The distribution of scaled revisits to any site before the death of the forager has been studied and the nature observed is exponential : $N(x)/S^{\mu}=N_0 \exp{-\lambda |x/S^{\nu}|}$. The scaling parameters $\mu$ and $\nu$ are same for very low $k/S$, but deviates slightly from each other as $k/S \rightarrow 1$, whereas the decay rate $\lambda \sim (k/S)^{-0.22}$. 

The energy of the forager and its food consumption are the other interesting quantities studied here. With increase of $S$, the energy decays slowly with time $t$ and the decay rate $\gamma \propto S^{-1.80}$. The decay rate is also slower for a higher $k/S$ and varies as $\gamma \propto (k/S)^{-0.24}$. 
The distribution of scaled food consumption over sites scales as $f(x)/S = F_0 \exp{-|x|/{\xi S^{\phi}}}$. The parameter $\phi$ decreases, whereas $\xi$ increases with increase of $k/S$. The total food consumption however varies exactly as lifetime $\tau$ and signifies that no step can be taken without food (fuel). More precisely, each food unit contributes one unit to the lifetime and survival for each time unit requires, on an average, the consumption of one unit of food. However, the consumption of food is not the only regulating factor for lifetime of the forager, and it has been observed that the strategy in terms of threshold energy and partial consumption affects the lifetime in a positive way. 

The collection of sites, either fully
or partially depleted of food, after the death of the forager  $N_{eat}$ shows a crossover behavior below and above $k/S \sim 0.5$. The behavior is seen to be not only dictated by the forager's starvation parameter $S$ and the threshold $k$, but also by the partial consumption. 

In the current study, although we have used a biologically
motivated minimalistic model, the same may be useful not only in understanding real-world foraging with complexities, but also for intermittent searching processes and for transport phenomena in disordered systems. Incorporation of intermittent resting, memory of food containing sites or intelligence to the model will affect its behavior and can be studied in future.

The authors thank Prof. Sidney Redner for some delightful discussions. MAM acknowledges financial support from CSIR, India (Grant no. 08/0463(12870)/2021-EMR-I). MAM and SG acknowledge
the computational facility of Vidyasagar College, University of Calcutta. 

\end{document}